\documentclass[a4paper,11pt]{article}
\pdfoutput=1

\usepackage{jcappub}

\usepackage{amssymb,journals,microtype,mymath,mathtools}
\usepackage[varg]{txfonts}

\title{Kinetic field theory: generic effects of alternative gravity theories on non-linear cosmic density-fluctuations}

\author[a]{A. Oestreicher,}
\author[b,c,d]{L. Capuano,}
\author[d]{S. Matarrese,}
\author[a,e]{L. Heisenberg}
\author[a,1]{and M. Bartelmann\note{Corresponding author.}}

\affiliation[a]{Institute for Theoretical Physics, Heidelberg University, \\ Philosophenweg 12, D-69120 Heidelberg, Germany}
\affiliation[b]{SISSA and INFN, Sezione di Trieste, \\
	Via Bonomea 265, I-34136 Trieste, Italy}
\affiliation[c]{IFPU - Institute for Fundamental Physics of the Universe, \\
	 Via Beirut 2, I-34014 Trieste, Italy}
\affiliation[d]{Department of Physics and Astronomy ``Galileo Galilei'', University of Padova, \\ Via Marzolo 8, I-35131 Padova, Italy}
\affiliation[e]{Institute for Theoretical Physics, ETH Zürich, \\Wolfgang-Pauli-Str. 27, SUI-8093 Zürich, Switzerland}

\emailAdd{oestreicher@thphys.uni-heidelberg.de}
\emailAdd{lcapuano@sissa.it}
\emailAdd{sabino.matarrese@pd.infn.it}
\emailAdd{l.heisenberg@thphys.uni-heidelberg.de}
\emailAdd{bartelmann@uni-heidelberg.de}

\abstract{
Non-linear cosmic structures contain valuable information on the expansion history of the background space-time, the nature of dark matter, and the gravitational interaction. The recently developed kinetic field theory of cosmic structure formation (KFT) allows to accurately calculate the non-linear power spectrum of cosmic density fluctuations up to wave numbers of $k\lesssim10\,h\,\mathrm{Mpc}^{-1}$ at redshift zero. Cosmology and gravity enter this calculation via two functions, viz.\ the background expansion function and possibly a time-dependent modification of the gravitational coupling strength.

The success of the cosmological standard model based on general relativity suggests that cosmological models in generalized theories of gravity should have observable effects differing only weakly from those in standard cosmology. Based on this assumption, we derive the functional, first-order Taylor expansion of the non-linear power spectrum of cosmic density fluctuations obtained from the mean-field approximation in KFT in terms of the expansion function and the gravitational coupling strength. This allows us to study non-linear power spectra expected in large classes of generalized gravity theories. To give one example, we apply our formalism to generalized Proca theories.
}

\begin{document}
\maketitle
\flushbottom

\section{Introduction}

Non-linear cosmic structures are mostly being studied with large-scale, well-resolved numerical simulations which have reached an impressive level of detail and sophistication \cite[for recent examples]{2021MNRAS.506.2871S, 2018MNRAS.475..676S}. Analytic methods have conventionally been based on the Euler-Poisson system of equations. Perturbation theory in either the Eulerian or Lagrangian picture, or effective field theories of cosmic structure formation, have also been developed to a formidable state \cite{2002PhR...367....1B, 2006PhRvD..73f3519C, 2012JCAP...12..013A, 2008PhRvD..77f3530M, 2013PhRvD..87h3522V, 2014JCAP...05..022P, 2014PhRvD..89d3521H, 2016JCAP...07..052B}. Due to limitations by the shell-crossing problem, they are confined to the weakly non-linear regime.

The shell-crossing problem is fundamentally caused by approximating dark matter as a fluid. It can be bypassed by dissolving the cosmic density field into particles whose trajectories through phase-space are described by the Hamiltonian equations. The statistical properties of the Hamiltonian flow of large particle ensembles can be encapsulated into a generating functional, integrating the transition probability of classical particles over a probability distribution of appropriately correlated initial phase-space positions. This is the fundamental idea of the kinetic field theory for cosmic structures (KFT, \cite{2016NJPh...18d3020B}), which we will briefly review in the next Section.

We could show in an earlier paper that the gravitational interactions between the KFT particles can be modelled in a mean-field approximation, similar to mean-field approaches in solid-state physics \cite{2021ScPP...10..153B}. We have derived a simple, analytic equation for the non-linear evolution of the density-fluctuation power spectrum whose two parameters can be estimated from within KFT itself, or mildly adapted to return values differing from those obtained from numerical simulations at the level of a few per cent down to Megaparsec scales at redshift zero.

This result encourages us to study in this paper possible effects on non-linear density-fluctuation power spectra in wide classes of modified gravity theories. The essential idea is as follows: A multitude of different types of observation tightly constrains the cosmological standard model founded upon general relativity. In generalized theories of gravity, changes in the expansion function $E(a)$ of the background space-time and possibly the gravitational coupling strength $G(a)$ relative to the cosmological standard model should be small. These functions are the only ingredients KFT needs from cosmology and from the theory of gravity. It should thus be permitted to evaluate the non-linear density-fluctuation power spectrum $P_\delta^\mathrm{(nl)}$ in a functional, first-order Taylor expansion in terms of these functions,
\begin{equation}
  \Delta P_\delta^\mathrm{(nl)}(k, a) = \int_{a_\mathrm{ini}}^a\D x\,\left[
    \frac{\delta P_\delta^\mathrm{(nl)}(k, a)}{\delta E(x)}\Delta E(x)+
    \frac{\delta P_\delta^\mathrm{(nl)}(k, a)}{\delta G(x)}\Delta G(x)
  \right]\;,
\label{eq:1}
\end{equation}
where $\Delta[E, G](x)$ are the differences between the respective functions in a generalized gravity theory relative to the standard cosmological model at scale factor $x$. The functional derivatives have to be taken within the standard cosmological model and can thus be calculated once and for all. We note that the non-linear density-fluctuation power spectrum in KFT also depends on the linear growth factor $D_+$ because this is a convenient time coordinate for KFT, but variations in $D_+$ are determined by those in $E$ and $G$ via the linear growth equation.

Besides applications to modified theories of gravity our formalism can also be applied to cosmological theories that keep general relativity as the underlying theory of gravity, but introduce non-trivial dark energy components such as dark energy with a time dependent equation of state. Such theories would suggest a modified expansion function $E(a)$ and can thus be implemented via the first term in \eqref{eq:1}.

We calculate in this paper the functional derivatives of the non-linear density-fluctuation power spectrum with respect to $E$ and $G$, starting from the mean-field approximation of KFT. These functional derivatives will be the main result of Sect.\ 3. In Sect. 4, we will illustrate the results at the example of one particular class of generalizations of general relativity, viz.\ the generalized Proca theories \cite{2014JCAP...05..015H}. In Sect.\ 2, we will begin by briefly reviewing the essential concepts of KFT and the mean-field equation for the non-linear power spectrum.

\section{The KFT mean field power spectrum}

Kinetic field theory (KFT) is a statistical theory for the evolution of classical particle ensembles in or out of equilibrium \cite{2012JSP...149..643D, 2016NJPh...18d3020B, 2019AnP...53100446B, 2022arXiv220211077K}. It defines an initial state of the ensemble by the probability distribution for phase-space positions $x^\mathrm{(i)}$ to be occupied. The Hamiltonian equations of motion for the particles on the expanding cosmological background allow constructing a retarded Green's function evolving the particle trajectories forward in time, including particle interactions. This Hamiltonian phase-space flow defines a diffeomorphic map of the initial phase-space distribution to any later time.

The information on the initial state of the particle ensemble and its time evolution are encapsulated in a generating functional $Z$. By functional derivatives of $Z$ with respect to suitably incorporated source fields, statistical information on the evolved ensemble can be extracted, such as the power spectrum or higher-order spectra of evolved density fluctuations. This approach based on particle trajectories in phase space has three major advantages compared to more conventional, analytic methods for studying cosmic structure formation. First, particle trajectories in phase space do not cross, avoiding by construction the shell-crossing problem notorious in cosmology. Second, with the Zel'dovich approximation or related approximations \cite{1970A&A.....5...84Z, 2015PhRvD..91h3524B}, a free reference motion or inertial motion can be chosen for the particles which already incorporates part of the gravitational interaction. Third, even small perturbations of particle trajectories can lead to arbitrarily high densities, which allows entering the regime of non-linear density evolution with low-order perturbation theory.

Particle interactions are described by an interaction operator acting on the generating functional $Z$. Taylor expansion of this operator opens a systematic approach to perturbation theory. Calculating the non-linear density-fluctuation power spectrum at first perturbative order returns a result which agrees with the spectrum obtained from numerical simulations within 10\ldots20~\% up to $k\lesssim10\,h\,\mathrm{Mpc}^{-1}$ at redshift zero \cite{2016NJPh...18d3020B}.

Even better agreement at the level of $\lesssim5$~\% between analytical and numerical, non-linear power spectra can be obtained by averaging the interaction operator in a mean-field approximation \cite{2021ScPP...10..153B}. In this approach, the non-linear power spectrum $P^\mathrm{(nl)}_\delta$ is related by
\begin{equation}
  P^\mathrm{(nl)}_\delta(k, t) \approx
  \E^{\langle S_I\rangle(k, t)}P^\mathrm{(lin)}_\delta(k, t)
\label{eq:2}
\end{equation}
to the linear power spectrum $P^\mathrm{(lin)}_\delta(k, t)$, where the scale- and time-dependent, mean-field averaged interaction term is
\begin{equation}
  \langle S_I\rangle(k, t) = 3\int_0^t\D t'\,
  \frac{a}{m}g_\mathrm{H}(t,t')D_+^2\sigma_J^2G\;, \quad
  g_\mathrm{H}(t,t') = \int_{t'}^t\frac{\D\bar t}{\bar m}\;.
\label{eq:3}
\end{equation}
Here, $D_+$ is the linear growth factor of cosmic density fluctuations, $g_{\mathrm{H}}$ is the so-called Hamilton propagator obtained by solving the equations of motion, $a$ is the scale factor, $G$ is the possibly time dependent gravitational coupling strength normalized at some initial time and $m$ is the effective particle mass
\begin{equation}
  m(t) = a^3(t)\frac{\D t}{\D a}E(t)\;,
\label{eq:4}
\end{equation}
where $E(t) = H(t)/H_\mathrm{i}$ is the cosmological expansion function, i.e.\ the Hubble function normalized by the Hubble constant at the same initial time. A suitable choice for this time is the time of cosmic recombination. For convenience, we also normalize the scale factor and the growth factor to unity at this initial time. The effective particle mass is thus unity initially and grows with time, reflecting the weakening gravitational interaction in an expanding space-time. \\ \\
The quantity $\sigma_J^2$ is a second moment of the damped initial density-fluctuation power spectrum,
\begin{equation}
  \sigma_J^2(k,t) = \frac{1}{(2\pi)^2}\int_0^\infty\,\d y
  y^2\tilde P^\mathrm{(i)}(y)J(y/k,y_0/k)\;,
\label{eq:5}
\end{equation}
effectively low-pass filtered by a filter function $J$. Since the damping depends on time, so does $\sigma_J^2$. The damped initial power spectrum is given by
\begin{equation}
  \tilde P^\mathrm{(i)}(y) = \left(1+Q_\mathrm{D}\right)^{-1}
  P^\mathrm{(i)}(y)\;,
\label{eq:6}
\end{equation}
where
\begin{equation}
  Q_\mathrm{D} = y^2\lambda^2(t)\;, \quad
  \lambda(t) \approx \frac{t}{1+\sqrt{t/\tau}}\sigma_1\; \quad \textrm{with} \quad \tau \approx 24.17.
\label{eq:7}
\end{equation}
$\lambda$ is a damping scale and $\sigma_1$ is a moment of the density-fluctuation power spectrum
\begin{equation}
    \sigma_n^2:= \frac{1}{2\pi^2}\int_0^{\infty}\d k k^{2n-2}P_{\delta}^{(i)}(k)\;.
\end{equation}
For a complete derivation of these expressions the reader is referred to \cite{2021ScPP...10..153B}. As explained in \cite{2021ScPP...10..153B} the power spectrum in the mean field approximation has two open parameters the non-linear scale $k_0$ and an effective viscosity $\nu$. These can either be estimated within KFT or chosen to optimize agreement with numerical results. For this paper we use the optimized parameters and leave them constant throughout the paper.

The time coordinate $t$ in (\ref{eq:3}) is set by the linear growth factor, $t = D_+-1$. While this is a suitable choice for deriving the formalism, our main goal in this paper is to ask how the power spectrum changes at a given scale factor (usually $a_0=1000$, i.e today). Rewriting the above expression with the scale factor as our independent variable and inserting the definition of the effective particle mass, the power spectrum takes the following simple form
\begin{equation}
  P^\mathrm{(nl)}_\delta(k, a) \approx
  \E^{\langle S_I\rangle(k, a)}P^\mathrm{(lin)}_\delta(k, a)
\label{eq:8}
\end{equation}
with the interaction term as follows
\begin{align}
  \langle S_I\rangle(k, a) = 3\int_{a_\mathrm{min}}^a\D a'\,
  g_\mathrm{H}(a,a')\frac{D_+^2\sigma_J^2G}{a'^2 E}\;, \quad
  g_\mathrm{H}(a,a') = \int_{a'}^a\frac{\D\bar a}{\bar a^3E}\;.
\label{eq:9}
\end{align}
The approximate equation (\ref{eq:8}) for the non-linear density-fluctuation power spectrum serves as the starting point for the present paper.

\section{Functional Taylor expansion of the non-linear density-fluctuation power spectrum}

The theory of gravity enters into the KFT formalism in two ways: (1) through the gravitational coupling strength $G$, which becomes time and possibly scale-dependent in some theories, and (2) through the time evolution of the background space-time in terms of the expansion function $E$. Here, we focus on theories that entail only a time dependence of $G$, but no dependence on scale. There are also classes of theories that introduce a screening mechanism reflected by a further scale dependence of $G$. We disregard such theories here to study different effects in isolation, but will accordingly generalize the formalism presented here in an upcoming publication \cite{Saxena2023}.

We know from observations that, if there are deviations from general relativity (GR), they have to be small. At least on small scales and for weak fields, any alternative theory of gravity (AG) would have to reproduce the predictions of GR. A successful alternative theory can thus only differ by a small amount from general relativity, and possible differences can only occur on cosmological scales or in strong fields. We can thus reasonably expect that such AG theories should predict only small corrections to the gravitational coupling and the expansion function.

Under this assumption, we can express the non-linear density-fluctuation power spectrum in an alternative theory of gravity as a first-order functional Taylor expansion around its form predicted by general relativity,
\begin{align}
  P_\delta^\mathrm{(nl)}[E+\delta E,G+\delta G](k,a) &\approx
  P_\delta^\mathrm{(nl)}[E,G] \nonumber\\ &+
  \int_{a_\mathrm{ini}}^a\D x\,
  \frac{\delta P_\delta^\mathrm{(nl)}[E,G]}{\delta E(x)}\Delta E(x)+
  \int_{a_\mathrm{ini}}^a\D x\,
  \frac{\delta P_\delta^\mathrm{(nl)}[E,G]}{\delta G(x)}\Delta G(x)\;.
\label{eq:10}
\end{align}
This expansion can be understood in the following way: The functional derivatives $\delta P_\delta^\mathrm{(nl)}/\delta E$ and $\delta P_\delta^\mathrm{(nl)}/\delta G$ quantify by how much the non-linear power spectrum $P_\delta^\mathrm{(nl)}$ evaluated at scale factor $a$ changes if the expansion function or the gravitational coupling strength are varied at some scale factor $a_{\mathrm{ini}}\le x\le a$ prior to $a$. We shall call $x$ the perturbance scale factor hereafter. The functions $\Delta E(x) = E_\mathrm{AG}(x)-E_\mathrm{GR}(x)$ and $\Delta G(x) = G_\mathrm{AG}(x)-G_\mathrm{GR}(x)$ are the changes in the expansion function and the gravitational coupling in the AG theory relative to GR. Since changes in $E$ and $G$ may occur at any scale factor preceding $a$, we need to integrate over the scale factor.

The functional derivatives are to be evaluated in GR and thus independent of the specific AG theory. The only terms that need to be specified for each theory are the changes $\Delta E(x)$ and $\Delta G(x)$. We are thus able to predict the non-linear power spectrum in the mean-field approximation of KFT for any alternative theory of gravity satisfying the assumption of small deviations from GR in the background expansion and the gravitational coupling.

\subsection{Functional derivatives of the growth factor}

For calculating the functional derivatives of the power spectrum with respect to $E$ and $G$, we need the functional derivatives of the linear growth factor with respect to both functions. This is because the most appropriate time coordinate for KFT is $t=D_+-1$, which introduces an indirect dependence of the power spectrum on $E$ and $G$ through this time $t$. To determine it, we take the functional derivatives of the linear growth equation
\begin{equation}
  D_+''(a)+\left(
    \frac{3}{a}+\frac{E'(a)}{E(a)}
  \right)D_+'-\frac{3}{2}\frac{\Omega_m(a)}{a^2}D_+ = 0
\label{eq:11}
\end{equation}
with respect to $E$ and $G$ and commute them with the derivative with respect to the scale factor $a$. KFT initial conditions are set at the time of recombination allowing us to neglect radiation when writing down \eqref{eq:11}. Additionally we assume dark energy perturbations to be negligible. The functional derivative of the growth factor with respect to the expansion function has already been determined in \cite{2020arXiv201103202S},
\begin{equation}
  \frac{\delta D_+(a)}{\delta E(x)} = \Theta(a-x)\,D_+(a)\left(
    f_E(x)\int_x^a\D y\,\frac{1}{y^3 D_+^2(y)E(y)}-\frac{D_+'(x)}{D_+(x)E(x)}
  \right)
\label{eq:12}
\end{equation}
with
\begin{equation}
  f_E(x) = x D_+^2(x)\left(\Omega_m^{2\gamma-1}-\frac{3}{2}\right)\Omega_m\;,
\label{eq:13}
\end{equation}
where the exponent $\gamma$ appears by writing the logarithmic derivative $\D\ln D_+/\D\ln a = \Omega_m^\gamma(a)$. For the $\Lambda$CDM reference model, $\gamma\approx 6/11$ \cite{2022ScPA....2....1H}.

\begin{figure}[ht]
  \includegraphics[width=0.49\hsize]{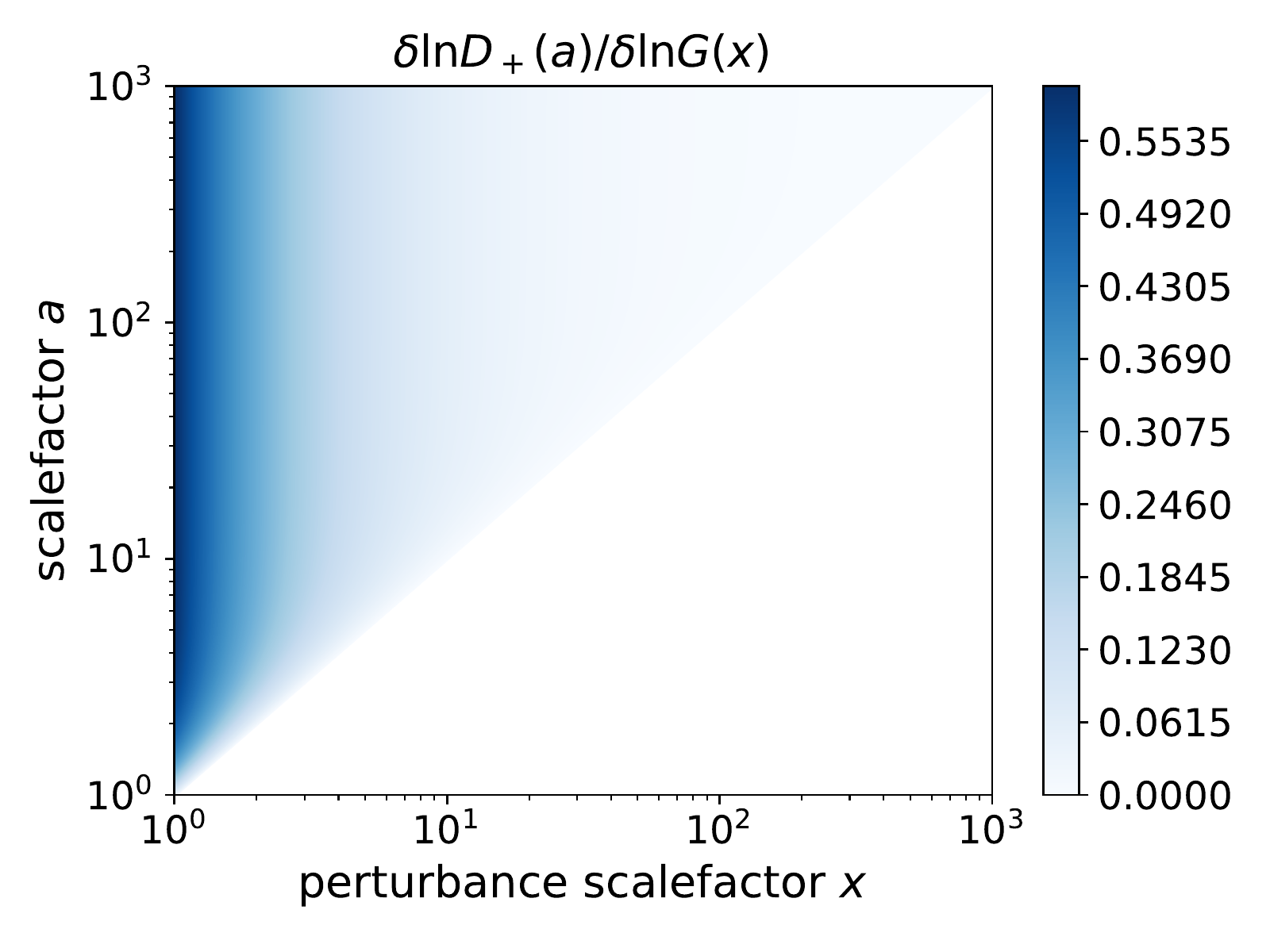}\hfill
  \includegraphics[width=0.49\hsize]{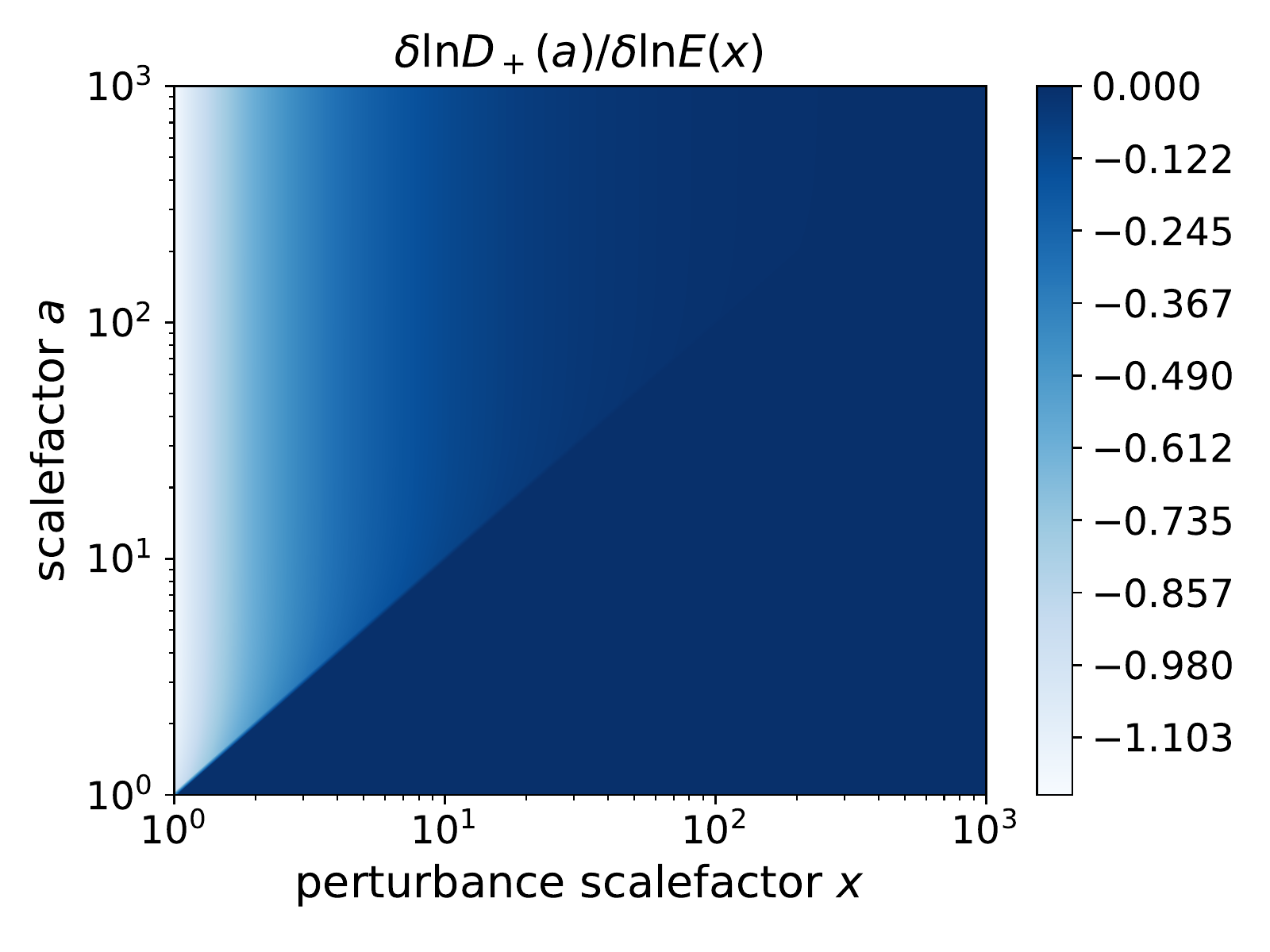}
\caption{Logarithmic functional derivatives of the linear growth factor $D_+$ with respect to the gravitational constant $G$ (\emph{left}) and the expansion function $E$ (\emph{right}) plotted as functions of the perturbance scale factor $x$ and the scale factor $a$.}
\label{fig:1}
\end{figure}

The derivative of the linear growth equation (\ref{eq:11}) with respect to the gravitational coupling, assuming that $\delta E/\delta G = 0$, is
\begin{equation}
  \frac{\D^2}{\D a^2}\frac{\delta D_+(a)}{\delta G(x)}+\left(
    \frac{3}{a}+\frac{E'(a)}{E(a)}
  \right)\frac{\D}{\D a}\frac{\delta D_+(a)}{\delta G(x)}-
  \frac{3}{2}\frac{\Omega_m(a)D_+(a)}{a^2}\left(
    \frac{\delta\ln D_+(a)}{\delta G(x)}+
    \frac{\delta\ln\Omega_m(a)}{\delta G(x)}
  \right) = 0\;.
\label{eq:14}
\end{equation}
Inserting the derivative
\begin{equation}
  \frac{\delta \Omega_m(a)}{\delta G(x)} =
  \frac{\delta}{\delta G(x)}\left(\frac{8\pi G(a)\rho_m}{3H^2(a)}\right) =
  \frac{\Omega_m(a)}{G(a)}\,\delta_\mathrm{D}(a-x)
\label{eq:15}
\end{equation}
into \eqref{eq:14} leads to the second-order inhomogeneous differential equation
\begin{equation}
  \frac{\D^2}{\D a^2}\frac{\delta D_+(a)}{\delta G(x)}+\left(
    \frac{3}{a}+\frac{E'(a)}{E(a)}
  \right)\frac{\D}{\D a}\frac{\delta D_+(a)}{\delta G(x)}-
  \frac{3}{2}\frac{\Omega_m(a)}{a^2}\frac{\delta D_+(a)}{\delta G(x)} =
  \frac{3}{2}\frac{D_+(a)}{a^2}\frac{\Omega_m(a)}{G(a)}\,
  \delta_\mathrm{D}(a-x)\;.
\label{eq:16}
\end{equation}
This differential equation for the function $\delta D_+/\delta G$ has the same shape as \eqref{eq:11}, which suggests the ansatz $\delta D_+(a)/\delta G(x) = C(a,x)D_+(a)$, which corresponds to the familiar variation of constants. Requiring further that our result be proportional to the step function $\Theta(a-x)$ to ensure causality results in
\begin{equation}
  \frac{\delta D_+(a)}{\delta G(x)} =
  \Theta(a-x)\,D_+(a)f_G(x)\int_x^a\frac{\D y}{D_+^2(y)y^3E(y)}
\label{eq:17}
\end{equation}
with
\begin{equation}
  f_G(x) = \frac{3}{2}\frac{\Omega_m(x)}{G(x)}D_+^2(x)xE(x)\;.
\label{eq:18}
\end{equation}
We show the logarithmic functional derivatives of the growth factor with respect to $E$ and $G$ in Fig.~\ref{fig:1}.

The absolute values of both derivatives are largest for small $x$ and decrease with increasing $x$. Both derivatives increase with the scale factor $a$. This behaviour confirms the expectation that the earlier deviations occur in $E$ and $G$, and the longer they act, the larger is their effect. The derivative with respect to the gravitational coupling is strictly positive, reflecting that an increase in the coupling strength always leads to enhanced structure formation. The  derivative with respect to the expansion function on the other hand is strictly negative because the background expansion slows down structure growth. This confirms the behaviour one would intuitively expect.

\subsection{Functional derivatives of the power spectrum}

Having obtained the functional dependence of the growth factor on the gravitational coupling and the expansion function, we can proceed to calculate the functional derivatives of the non-linear power spectrum with respect to $G$ and $E$. They separate into derivatives of the linear power spectrum $P_\delta^\mathrm{(lin)}$ and of the mean interaction term $\langle S_I\rangle$. The functional derivatives with respect to $X = (E,G)$ are
\begin{equation}
  \frac{\delta P_\delta^\mathrm{(nl)}(k,a)}{\delta X(x)} =
  \E^{\langle S_I\rangle}\left(
    \frac{\delta P_\delta^\mathrm{(lin)}(k,a)}{\delta X(x)}+
    P_\delta^\mathrm{(lin)}(k,a)\frac{\delta\langle S_I\rangle(k,a)}{\delta X(x)}
  \right)\;.
\label{eq:19}
\end{equation}
We shall now calculate the functional derivatives of the linear power spectrum and of the interaction term in the next two sections, respectively.

\subsubsection{Derivatives of the linear power spectrum}

The linear power spectrum depends on $E$ and $G$ through the growth factor $D_+$, since its time evolution is given by the square of $D_+$. Its derivative with respect to $X$ is
\begin{align}
  \frac{\delta P_\delta^\mathrm{(lin)}(k,a)}{\delta X(x)} &=
  \frac{\delta}{\delta X(x)}\left(D_+^2(a)P_\delta^\mathrm{(i)}(k)\right)
  \nonumber \\ &=
  2D_+(a)P_\delta^\mathrm{(i)}(k)\frac{\delta D_+(a)}{\delta X(x)} =
  2P_\delta^\mathrm{(lin)}(k,a)\frac{\delta\ln D_+(a)}{\delta X(x)}\;.
\label{eq:20}
\end{align}
Equation (\ref{eq:20}) shows that these functional derivatives reproduce the shape of the linearly evolved power spectrum but change its amplitude by an amount proportional to the functional derivatives of the growth factor with respect to $G$ and $E$ shown in Fig. \ref{fig:1}.

\subsubsection{Derivatives of the interaction term}

The mean interaction term $\langle S_I\rangle$ depends both explicitly on the gravitational coupling and on the expansion function, and also implicitly on $G$ and $E$ through the growth factor $D_+$. We take the mean-field interaction term in the form (\ref{eq:9}). Taking its functional derivative with respect to $X=(G,E)$ and applying the product rule results in
\begin{align}
  \frac{\delta\langle S_I\rangle(k,a)}{\delta X(x)} &=
  3\int_{a_\mathrm{min}}^a\D a'\,g_\mathrm{H}(a,a')\frac{D_+^2\sigma_J^2G}{a'^2 E}
  \nonumber \\ &\cdot \left[
    \frac{\delta\ln g_\mathrm{H}(a,a')}{\delta X(x)}-
    \frac{\delta\ln E(a')}{\delta X(x)}+
    \frac{\delta\ln\sigma_J^2(a')}{\delta X(x)}+
    2\frac{\delta\ln D_+(a')}{\delta X(x)}+
    \frac{\delta\ln G(a')}{\delta X(x)}
  \right]\;.
\label{eq:21}
\end{align}
Any function appearing here is understood to depend on the integration variable unless otherwise specified. To proceed further we now need the functional derivatives of $D_+$, $g_{\mathrm{H}}$, and $\sigma_J^2$. Having already calculated the derivatives of the growth factor, we begin with the functional derivative of $g_{\mathrm{H}}(a,a')$. The Hamilton propagator only depends on the expansion function $E$ and not on the effective gravitational constant $G$. Pulling the derivative into the integral, we find
\begin{equation}
  \frac{\delta g_\mathrm{H}(a,a')}{\delta E(x)} =
  -\int^a_{a'}\frac{\D\bar a}{\bar a^3 E^2(\bar a)}\delta_\mathrm{D}(\bar a-x) =
  -\frac{\Theta(a-x)\Theta(x-a')}{x^3E^2(x)}\;.
\label{eq:22}
\end{equation}

Lastly we need the functional derivative of $\sigma_J^2$. It is neither a direct function of $E$ or $G$ and only depends on $E,G$ through our time variable $t=D_+-1$. The functional derivative thus is
\begin{align}
  \frac{\delta\sigma_J^2(a)}{\delta X(x)} =
  \dot\sigma_J^2\frac{\delta D_+(a)}{\delta X(x)}\;,
\label{eq:24}
\end{align}
where $\dot\sigma_J^2$ is the time derivative of $\sigma_J^2$. It can quickly be calculated using \eqref{eq:6}, resulting in
\begin{equation}
  \dot\sigma_J^2 = -\frac{1}{(2\pi)^2}
  \int_0^\infty\D y\,y^4\sigma_\nu^2\lambda^2(1+Q_\mathrm{D})^{-2}\left(
    \frac{2}{t}-\frac{1}{\sqrt{t\tau}+t}
  \right)\,P^\mathrm{(i)}(y)\,J(y/k,y_0/k)\;.
\label{eq:25}
\end{equation}

\begin{figure}[ht]
  \includegraphics[width=0.49\hsize]{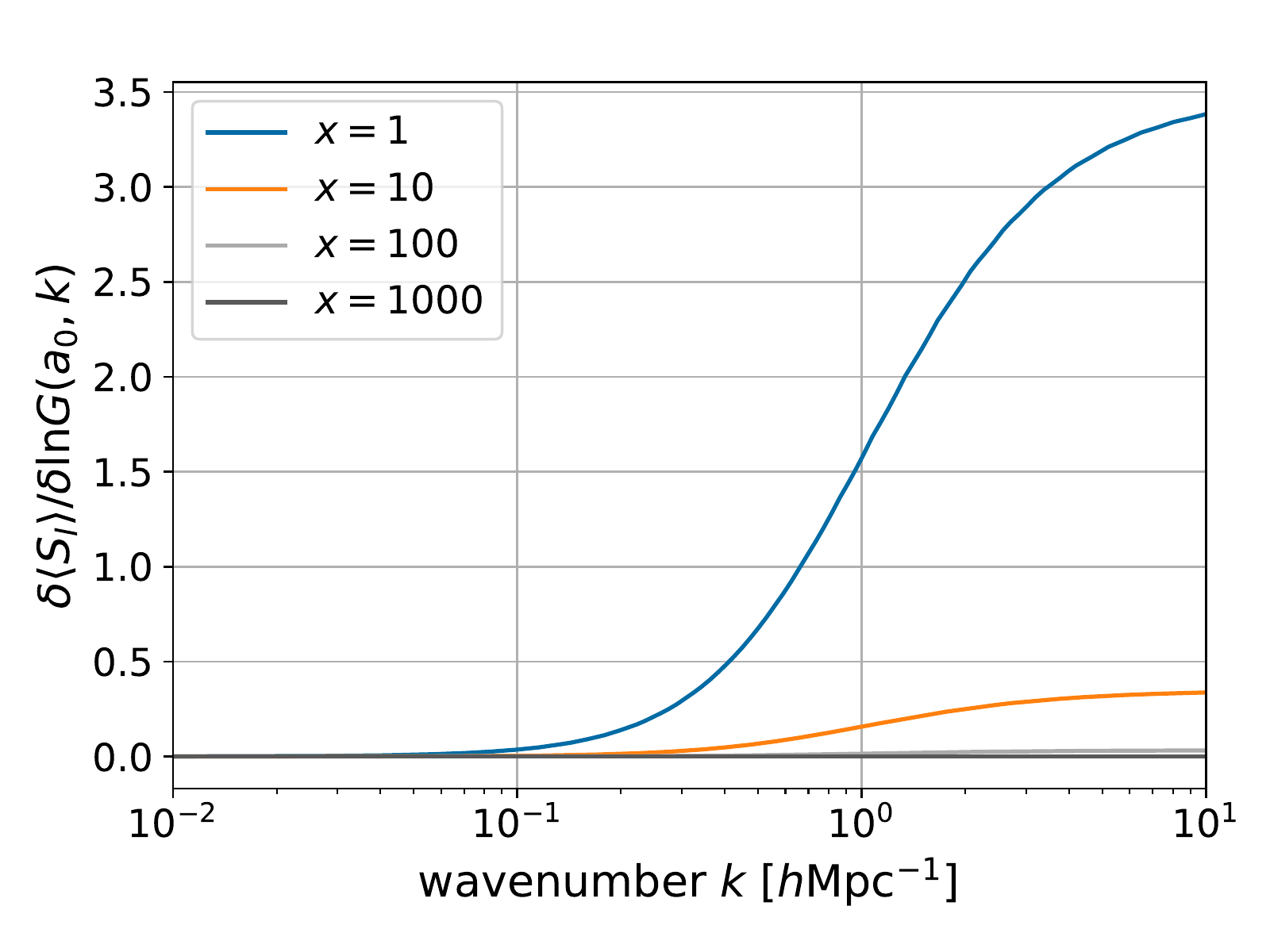}\hfill
  \includegraphics[width=0.49\hsize]{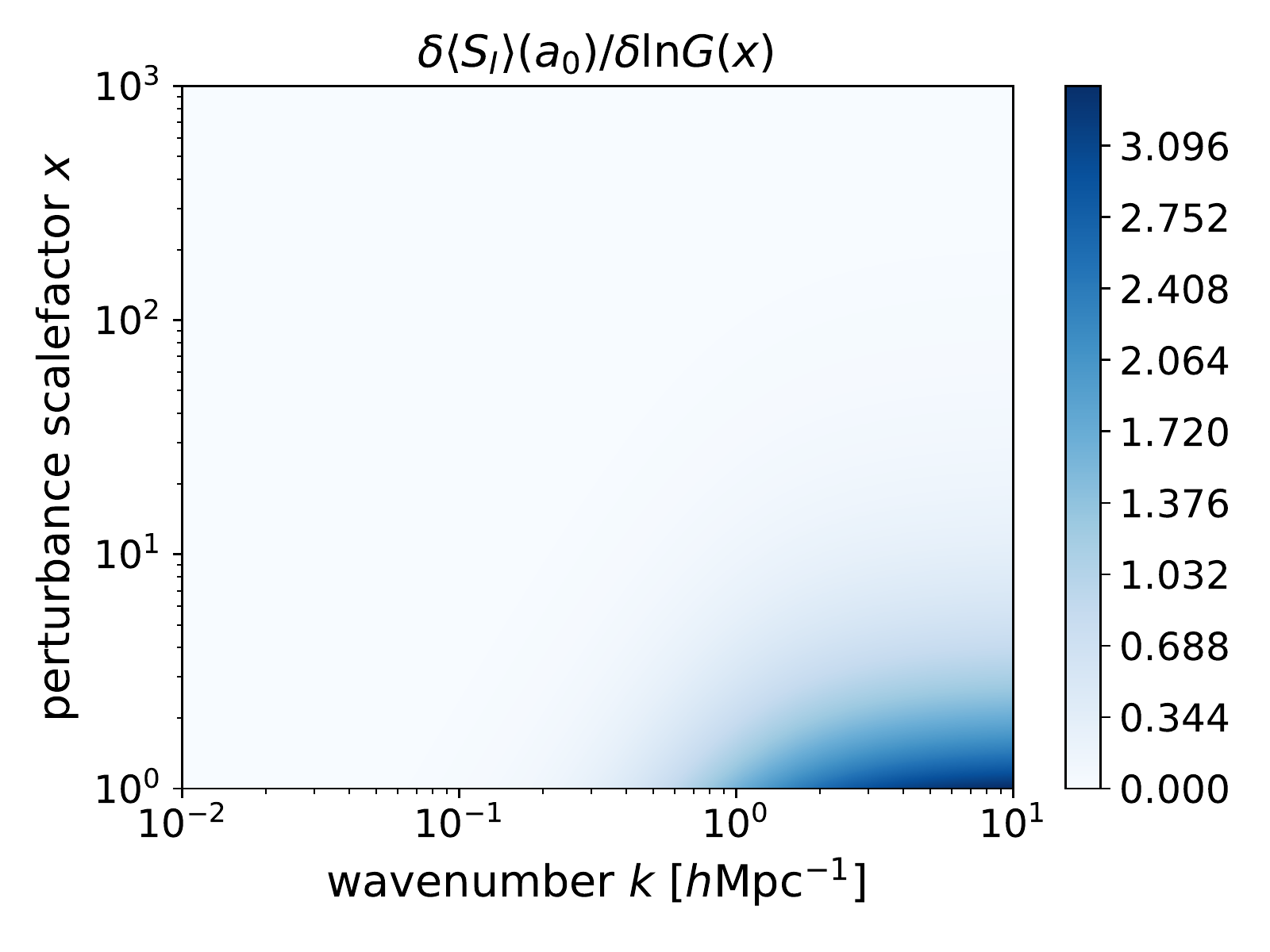}
\caption{Functional derivative of the mean interaction term $\langle S_I\rangle$ with respect to the logarithm of $G$ evaluated today at scale factor $a_0$, plotted as a function of the wavenumber $k$ and for different values of the perturbance scale factor $x$ (left), and plotted as a function of both $k$ and $x$ in a two-dimensional figure (right).}
\label{fig:2}
\end{figure}

\begin{figure}[ht]
  \includegraphics[width=0.49\hsize]{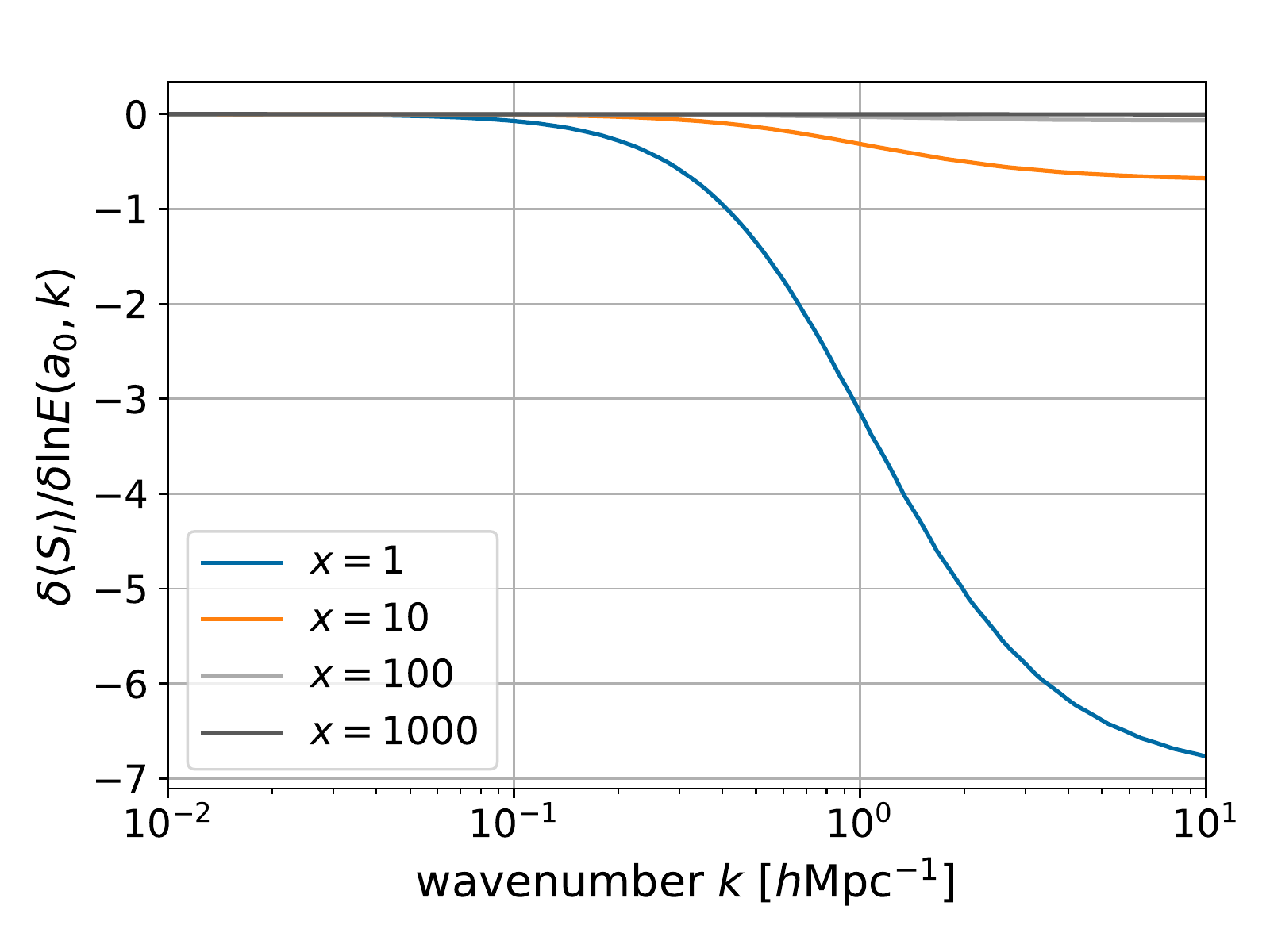}\hfill
  \includegraphics[width=0.49\hsize]{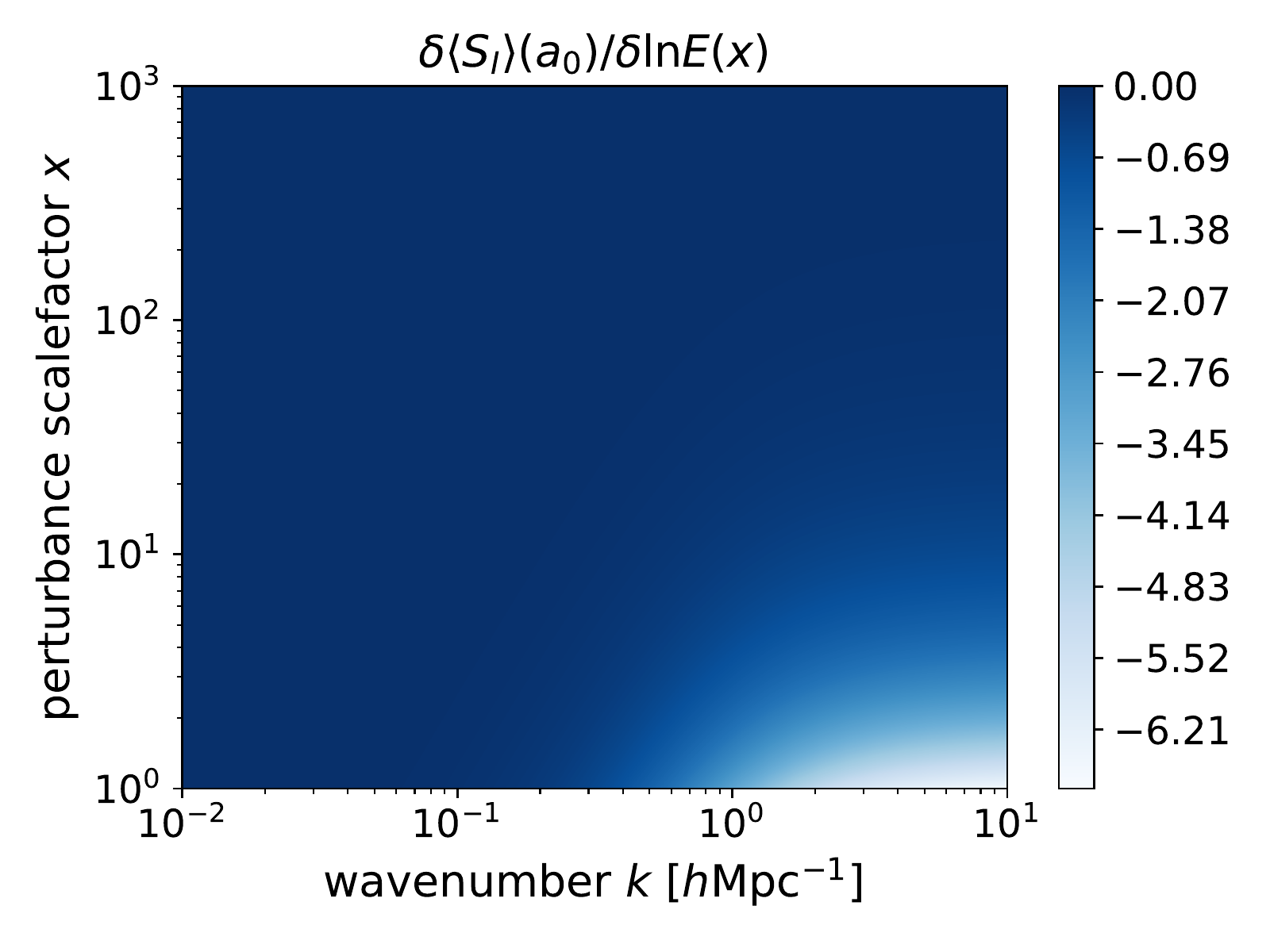}
\caption{As Fig.~\ref{fig:2} for the expansion function $E$ instead of the gravitational coupling $G$.}
\label{fig:3}
\end{figure}

Having calculated the functional derivatives of all functions appearing in the interaction term, we can now return to expression \eqref{eq:21}. The functional derivative of the expansion function and the Hamilton propagator vanish for $X=G$, while the derivative of $G$ returns a delta function. Integrating over it leaves us with
\begin{align}
  \frac{\delta\langle S_I\rangle(k,a)}{\delta G(x)} &=
  3\Theta(a-x)g_\mathrm{H}(a,x)\frac{D_+^2(x)\sigma_J^2(x)}{x^2 E(x)}
  \nonumber \\ &+
  \int_{a_\mathrm{min}}^a\D a'\,g_\mathrm{H}(a,a')
  \frac{D_+^2\sigma_J^2G}{a'^2 E}\left[
    \frac{\delta\ln\sigma_J^2(a')}{\delta G(x)}+
    2\frac{\delta\ln D_+}{\delta G(x)}
  \right]\;.
\label{eq:26}
\end{align}
Taking the functional derivative of the interaction term with respect to $X=E$, the functional derivative of the gravitational coupling vanishes, and we now get a delta function for the functional derivative of $E$. Integrating over it results in
\begin{align}
  \frac{\delta\langle S_I\rangle(k,a)}{\delta E(x)} &=
  -3\Theta(a-x)g_\mathrm{H}(a,x)\frac{D_+^2(x)\sigma_J^2(x)G(x)}{x^2 E^2(x)}
  \nonumber \\ &+
  3\int_{a_\mathrm{min}}^a\D a'\,g_\mathrm{H}(a,a')\frac{D_+^2\sigma_J^2G}{a'^2 E}
  \left[
    \frac{\delta\ln g_\mathrm{H}(a,a')}{\delta E(x)}+
    \frac{\delta\ln\sigma_J^2(a')}{\delta E(x)}+
    2\frac{\delta\ln D_+}{\delta E(x)}
  \right]\;.
\label{eq:27}
\end{align}
We show the functional derivatives of $\langle S_I\rangle$ with respect to $G$ and $E$ in Figs.~\ref{fig:2} and \ref{fig:3}. The mean interaction term remains unchanged on the largest scales since its functional derivative is non-zero only on moderate and small scales. Note that both derivatives are proportional to Heaviside functions, ensuring causality. As before, the derivative with respect to $G$ is strictly positive reflecting that an increase in the coupling strength enhances structure formation and the derivative with respect to $E$ is strictly negative reflecting that the background expansion slows down structure growth.

\subsubsection{Functional derivatives of the non-linear power spectrum}

\begin{figure}[ht]
  \includegraphics[width=0.49\hsize]{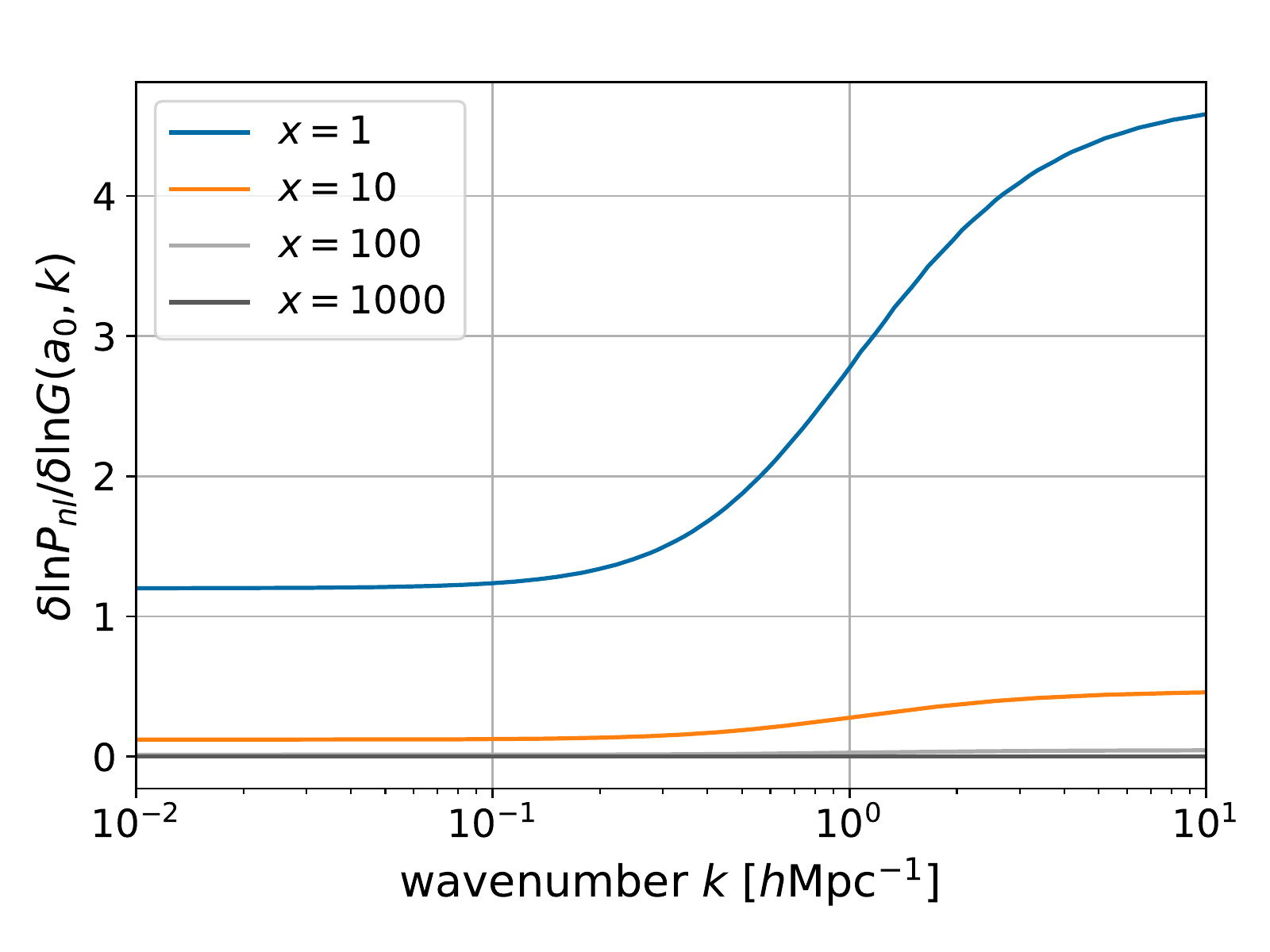}\hfill
  \includegraphics[width=0.49\hsize]{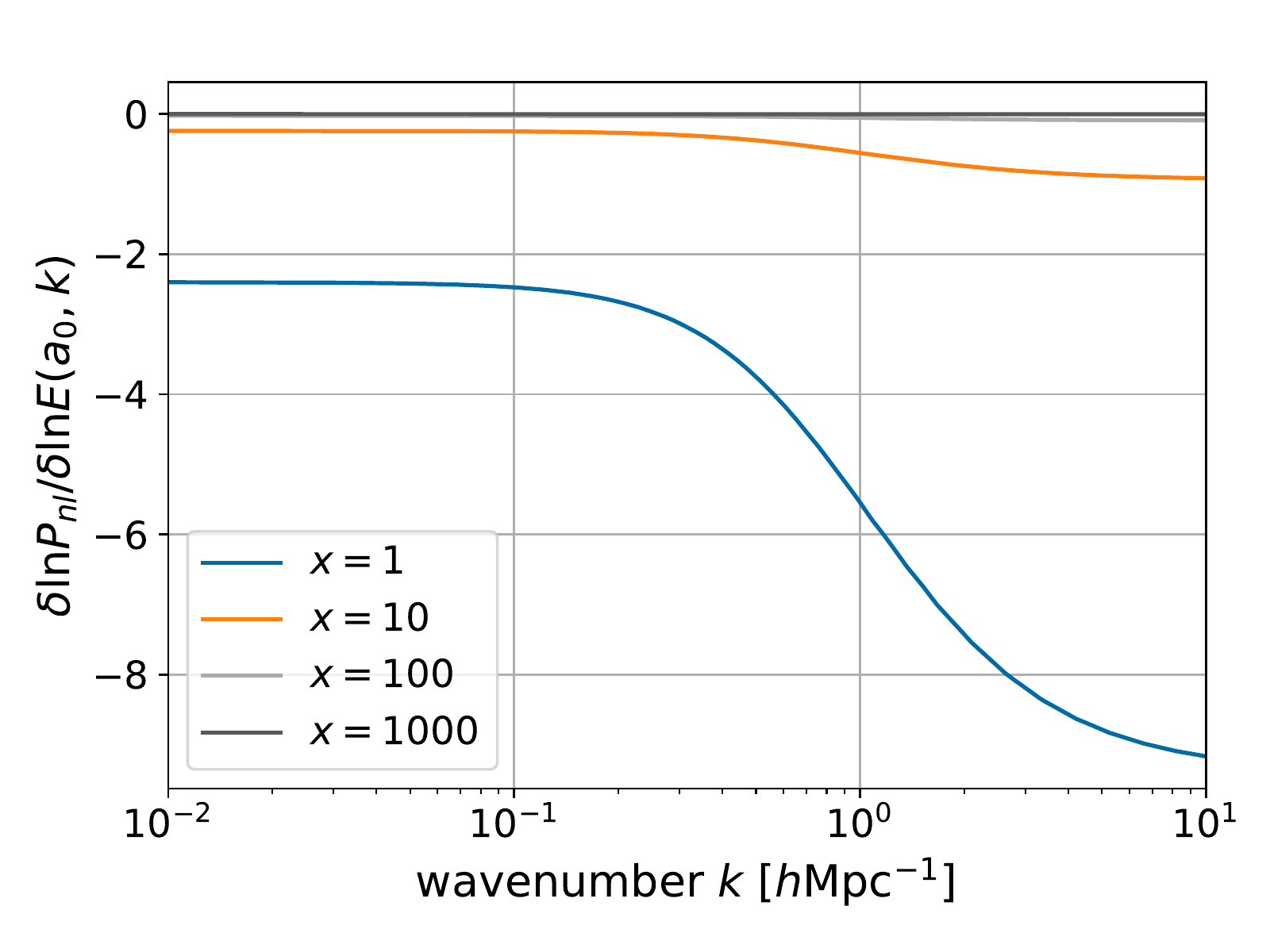}
\caption{Logarithmic functional derivatives of the non-linear power spectrum $P$ with respect to the gravitational constant $G$ (\emph{left}) and the expansion function $E$ (\emph{right}), evaluated today at scale factor $a_0$ and plotted for different perturbance scale factors $x$.}
\label{fig:4}
\end{figure}

Having calculated the functional derivatives of the linearly evolved power spectrum and of the mean interaction term, we can now return to the expression \eqref{eq:19} for the complete functional derivative of the non-linear power spectrum. The first term in both expressions describes a pure amplitude change of the power spectrum independent of $k$, while the second term does depend on $k$ and thus changes the shape of the non-linear relative to the linear power spectrum. We illustrate this in Fig.~\ref{fig:4}.

At large scales, the relative change of the power spectrum with respect to both $G$ and $E$ is constant, while it depends on $k$ at smaller scales. The $k$-independent amplitude change results from the functional derivatives of the linear power spectrum and appears because an alternative gravity theory may imply changes to the linear growth factor. Since the linear power spectrum grows as $D_+^2(a)$, the modified growth factor would then also change the overall amplitude of the power spectrum. The $k$ dependence on small scales results purely from the derivative of the mean interaction term. This part changes the shape of the non-linear power spectrum in response to modified gravity.

\subsection{Non-linear power spectra in alternative gravity theories}

With all necessary functional derivatives at hand, we can now write down the full expression for an alternative power spectrum as given by our functional first-order Taylor expansion
\begin{align}
  P_{\delta, \mathrm{AG}}^\mathrm{(nl)}(k,a) &=
  P_{\delta, \mathrm{GR}}^\mathrm{(nl)}(k,a) \nonumber\\ &\cdot
  \left[1+
    2\sum_{X=E,G}\int_{a_\mathrm{ini}}^a\D x
    \frac{\delta\ln D_+(a)}{\delta X(x)}\Delta X(x)+
    \sum_{X=E,G}\int_{a_\mathrm{ini}}^a\D x
    \frac{\delta\langle S_I\rangle(k,a)}{\delta X(x)}\Delta X(x)
  \right]\;.
\label{eq:28}
\end{align}
As discussed above, the derivatives predict two different types of change: (1) A $k$-independent, pure amplitude change (first integral in square brackets) and (2) a $k$-dependent and thus shape-changing part (second integral in square brackets).

Thus far, by construction of the Taylor expansion, the power spectrum for an alternative theory of gravity and general relativity are equal at the initial scale factor since all integrals in \eqref{eq:28} vanish there. For comparison with observations, we should rather set the amplitudes of the two spectra equal at the final scale factor $a_0$. Normalizing the spectra in this way, we only need to keep the shape-changing part. For the two spectra to reach the same amplitude today, their initial amplitudes must have differed, in order to account for their different time evolution. This difference in initial amplitude will also affect the interaction term \eqref{eq:9}, since it contains the moment $\sigma_J^2$, which also depends on the initial power spectrum. Assuming the amplitude change to be small, we can include it into our Taylor-expansion approach by adding another term
\begin{align}
  P_{\delta,\mathrm{AG}}^\mathrm{(nl)} \approx
  P_{\delta,\mathrm{GR}}^\mathrm{(nl)}+\sum_{X=E,G}
  \int_{a_{\mathrm{ini}}}^{a}\D x\,
  \frac{\delta P_{\delta,\mathrm{GR}}^\mathrm{(nl)}}{\delta X(x)}\Delta X(x)
  +\frac{\partial P_{\delta,\mathrm{GR}}^\mathrm{(nl)}}{\partial\mathcal{A}}\Delta\mathcal{A}\;,
\label{eq:29}
\end{align}
where $\Delta \mathcal{A}=\mathcal{A}_{\mathrm{AG}}-\mathcal{A}_{\mathrm{GR}}$ is the change in the amplitude of the initial power spectrum. We now evaluate this additional term.

The amplitude changes with the square of the growth factor. To correct it, we thus need to determine by how much the growth factor differs in the alternative theory. We do this by the first-order Taylor approximation
\begin{equation}
  D_{+, \mathrm{AG}}^2(a) = D_+^2(a)\left(
    1+\frac{2}{D_+(a)}\sum_{X=E,G}\int_{a_\mathrm{min}}^a\,\D x
    \frac{\delta D_+(a)}{\delta X(x)}\Delta X(x)
  \right)\;.
\label{eq:30}
\end{equation}
For the two spectra to reach the same amplitude today, the initial power spectrum for the AG theory should thus be lower by the factor
\begin{align}
  \mathcal{A}_\mathrm{AG} &= \left(
    1+\frac{2}{D_+(a)}\sum_{X=E,G}\int_{a_\mathrm{min}}^a\,\D x
    \frac{\delta D_+(a)}{\delta X(x)}\Delta X(x)
  \right)^{-1} \nonumber \\ &\approx
  1-\frac{2}{D_+(a)}\sum_{X=E,G}\int_{a_\mathrm{min}}^a\,\D x
  \frac{\delta D_+(a)}{\delta X(x)}\Delta X(x)\;.
\label{eq:31}
\end{align}
Setting $\mathcal{A}_{\mathrm{GR}}=1$ for the reference model, we thus find
\begin{align}
  \Delta\mathcal{A} \approx -\frac{2}{D_+(a)}\sum_{X=E,G}\int_{a_\mathrm{min}}^a\,\D x
  \frac{\delta D_+(a)}{\delta X(x)}\Delta X(x)\;.
\label{eq:32}
\end{align}
To calculate the derivative of the power spectrum with respect to $\mathcal{A}$, we introduce an explicit amplitude factor $\mathcal{A}$ into
\begin{equation}
  P^\mathrm{(nl)}_\delta(k, t) \approx
  \E^{\langle S_I\rangle(k, t)}D_+^2\mathcal{A}P^\mathrm{(i)}
\label{eq:33}
\end{equation}
and the function
\begin{equation}
  \sigma_J^2(k,t) = \frac{1}{(2\pi)^2}
  \int_0^\infty\D y\,y^2\left(1+\mathcal{A}Q_\mathrm{D}\right)^{-1}
  \mathcal{A}P^\mathrm{(i)}(y)J(y/k,y_0/k)
\label{eq:35}
\end{equation}
appearing in the mean-field interaction term $\langle S_\mathrm{I}\rangle$. The derivative of the power spectrum is
\begin{equation}
  \frac{\partial P^\mathrm{(nl)}_\delta}{\partial\mathcal{A}} =
  \frac{P^\mathrm{(nl)}_\delta}{\mathcal{A}}+
  P^\mathrm{(nl)}_\delta\,
  \frac{\partial\langle S_I\rangle(k,t)}{\partial\mathcal{A}}\;,
\label{eq:36}
\end{equation}
where
\begin{equation}
  \frac{\partial\langle S_I\rangle(k,t)}{\partial\mathcal{A}} =
  3\int_0^t\D t'\,g_\mathrm{H}(t,t')\frac{a}{m}D_+^2G
  \frac{\partial\sigma_J^2\left(k, t'\right)}{\partial\mathcal{A}}\;,
\label{eq:37}
\end{equation}
and
\begin{equation}
  \frac{\partial\sigma_J^2(k,t)}{\partial\mathcal{A}} =
  \frac{1}{(2\pi)^2}\int_0^\infty\,\D y y^2
  \left(1+\mathcal{A}Q_\mathrm{D}\right)^{-1}
  \mathcal{A}P^\mathrm{(i)}(y)\left[
    \frac{1}{\mathcal{A}}-Q_\mathrm{D}\left(1+\mathcal{A}Q_\mathrm{D}\right)^{-1}
  \right]J(y/k,y_0/k)\;.
\label{eq:38}
\end{equation}
Inserting the result for $\Delta \mathcal{A}$ into \eqref{eq:29} we can summarize our result as follows
\begin{align}
  P_{\delta,\mathrm{AG}}^\mathrm{(nl)}(k,a) &\approx
  P_{\delta,\mathrm{GR}}^\mathrm{(nl)}(k,a)+\sum_{X=E,G}
  \int_{a_\mathrm{ini}}^a\D x\,\left(
    \frac{\delta P_{\delta,\mathrm{GR}}^\mathrm{(nl)}(k,a)}{\delta X(x)}-
    2\frac{\partial P_{\delta,\mathrm{GR}}^\mathrm{(nl)}(k,a)}{\partial\mathcal{A}}
    \frac{\delta\ln D_+(a)}{\delta X(x)}
  \right)\Delta X(x) \nonumber \\ &\approx
  P_{\delta,\mathrm{GR}}^\mathrm{(nl)}(k,a)+\sum_{X=E,G}
  \int_{a_\mathrm{ini}}^{a}\D x\,\left(
    \frac{\delta P_{\delta,\mathrm{GR}}^\mathrm{(nl)}(k,a)}{\delta X(x)}+
    \mathcal{A}_X(k,a,x)
  \right)\Delta X(x)\;.
\label{eq:39}
\end{align}
We can now study a wide range of alternative power spectra, simply by specifying the change in the gravitational coupling $\Delta G$ and the change in the expansion function $\Delta E$ suggested by alternative gravity theories.

\subsection{Code and Data Release}

One major advantage of our approach is that the coefficients of the functional Taylor expansion are to be evaluated in the standard $\Lambda$CDM cosmological model and can thus be calculated once and for all. Together with this paper, we provide tables of the $\Lambda$CDM power spectrum and the coefficients under the integral in \eqref{eq:39} evaluated at scale factor $a_0=1000$, i.e.\ today. We also provide a simple Python script requiring an alternative expansion function and effective gravitational constant as input, performing the remaining integration in \eqref{eq:39} and returning the non-linear power spectrum. This should allow to quickly calculate the non-linear power spectrum for a desired alternative theory of gravity. The code can be found \href{https://github.com/oestreichera/kftAGrav}{here}.

\section{Application to Generalized Proca theories}

As a demonstration, we will now apply our approach to one specific alternative gravity theory, viz.~the generalized Proca theories \cite{2014JCAP...05..015H, 2016JCAP...06..048D}. We choose this model since KFT has already been applied to it in an earlier paper \cite{2019PhLB..796...59H}, albeit with another normalization of the interaction term and the final power spectrum.

General Relativity is the field theory which describes the dynamics of the metric tensor, whose equations of motion are the well known Einstein equations. In four dimensions, GR is the unique theory which can be constructed from an action only containing the metric tensor and its first and second derivative (Lovelock's theorem).

One possibility for extending GR is to include additional fields into the Einstein-Hilbert action. An important class of these theories is represented by vector-tensor theories, in which an additional vector field is included. Within this family, requiring second-order equations of motion, the most general theories are the so called generalized Proca theories. We now proceed to apply our functional Taylor expansion to this theory, with the model described in \cite{2016PhRvD..94d4024D} for the gravitational coupling.

In this case, the dark energy density parameter reads
\begin{equation}
  \Omega_\mathrm{DE} = 1-\Omega_\mathrm{m} =
  \frac{6p_2^2(2p+2p_2-1)\beta_4-p_2(p+p_2)(1+4p_2\beta_5)}{p_2(p+p_2)}\,y\;,
\label{eq:40}
\end{equation}
with $p$, $p_2$, $\beta_4$, $\beta_5$ and $y$ parameters of the theory. The equation-of-state parameter for dark energy itself depends on $\Omega_\mathrm{DE}$ as
\begin{equation}
  w_\mathrm{DE} = -\frac{1+s}{1+s\Omega_\mathrm{DE}}\;,
\label{eq:41}
\end{equation}
with $s = p_2/p$. The effective gravitational constant for the examined model reads
\begin{equation}
  \frac{G_\mathrm{eff}}{G} = \frac{(p+p_2)}{\mathcal{F}_G}\left\{
    q_Vu^2-2p_2y[1-6\beta_4(2p+2p_2-3)+2\beta_5(3p+2p_2-3)]
  \right\}\;,
\label{eq:42}
\end{equation}
where $q_V$ controls the amplitude of the correction to GR due to the introduction of the vector field, and $\mathcal{F}_G$ is a function of the other parameters of the model; see \cite{2016PhRvD..94d4024D} for the full expression.

Equipped with the cosmic expansion function and the effective gravitational coupling for this model, we are now able to compute the variations $\Delta E$ and $\Delta G$ for it and calculate the alternative power spectrum using our Taylor expansion.

\begin{figure}[ht]
  \includegraphics[width=0.49\hsize]{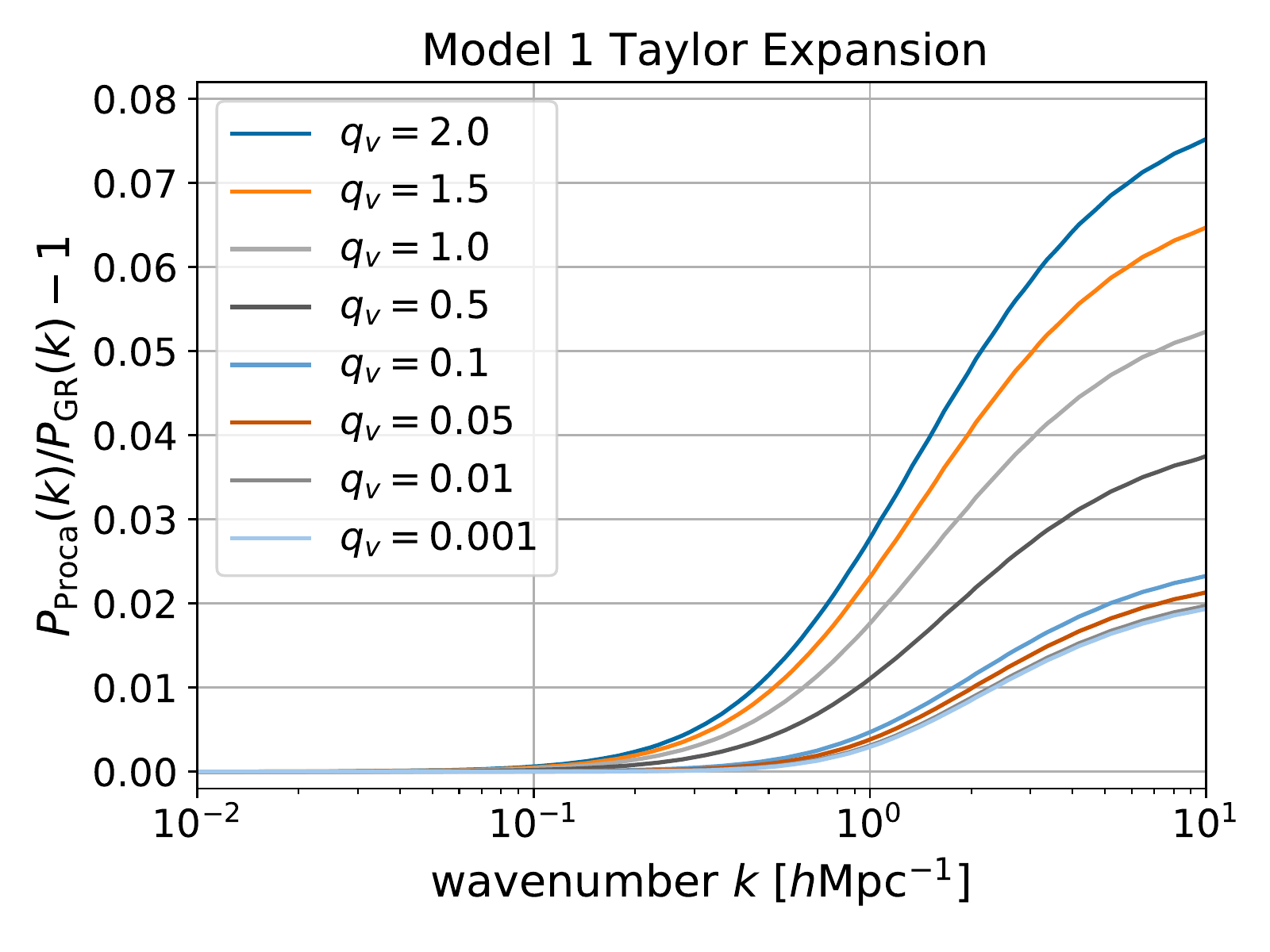}\hfill
  \includegraphics[width=0.49\hsize]{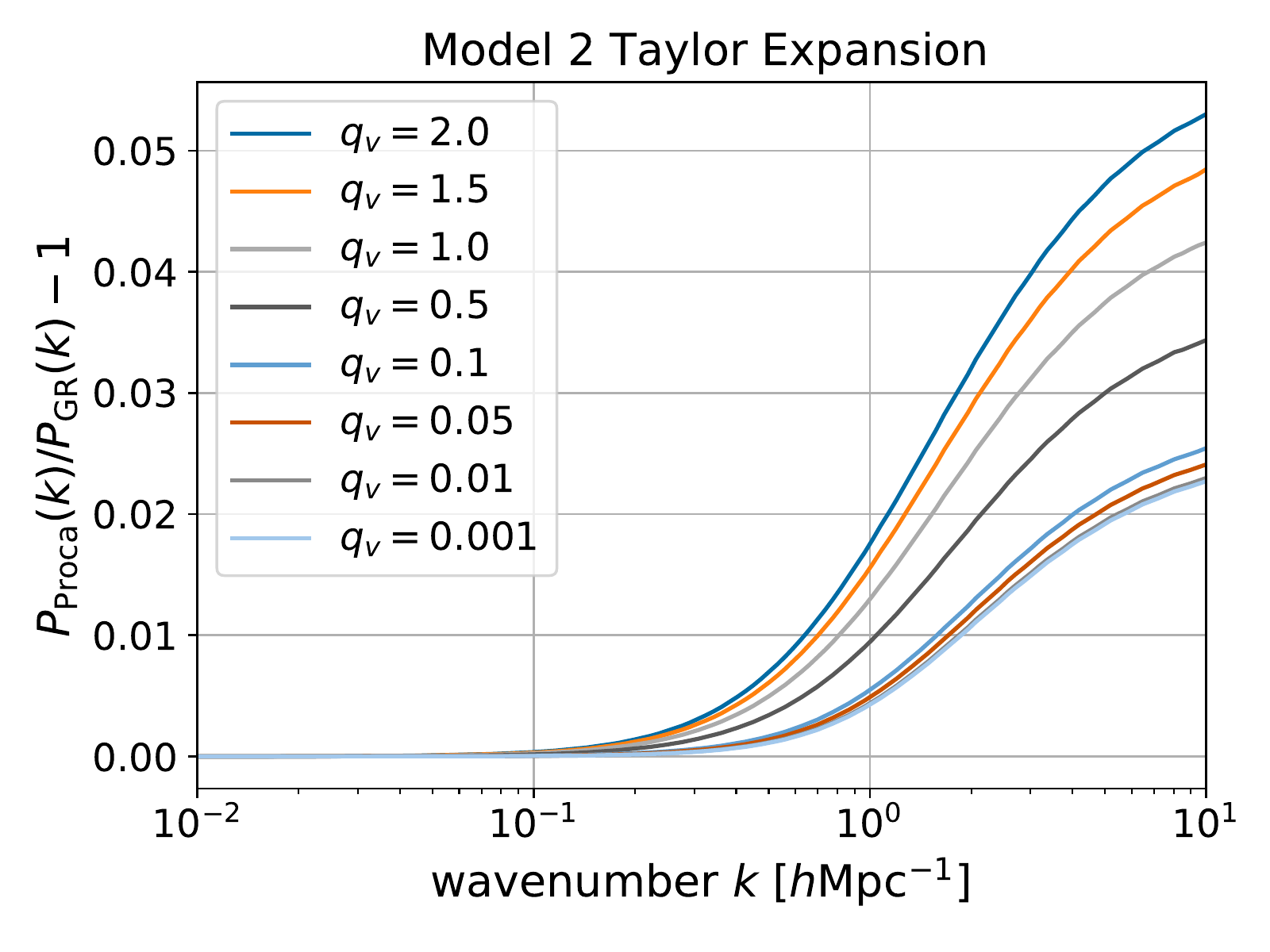}
\caption{Relative difference between the non-linear power spectrum for the generalized Proca theories and the power spectrum for the standard cosmological model, for two different Proca models. Model~1 (\emph{left}) with the parameter choices  $p=2.5, p_2=0.5, \lambda=0.86, \beta_4=1.0e-4, \beta_5=0.052$ and Model~2 (\emph{right}) with $p=2.5, p_2=0.5, \lambda=0.86, \beta_4=0.0, \beta_5=0.0$, both plotted for a range of values for $q_v$. Results obtained with the Taylor expansion.}
\label{fig:5}
\end{figure}

As in \cite{2019PhLB..796...59H}, we implement two different models with the parameter choices $p=2.5, p_2=0.5, \lambda=0.86, \beta_4=1.0e-4, \beta_5=0.052$ and $p=2.5, p_2=0.5, \lambda=0.86, \beta_4=0.0, \beta_5=0.0$, respectively. For both models, we plot results for a range of values for $q_v$. The result obtained with our Taylor expansion \eqref{eq:29} can be seen in Fig.~\ref{fig:5}.

\begin{figure}[ht]
  \includegraphics[width=0.49\hsize]{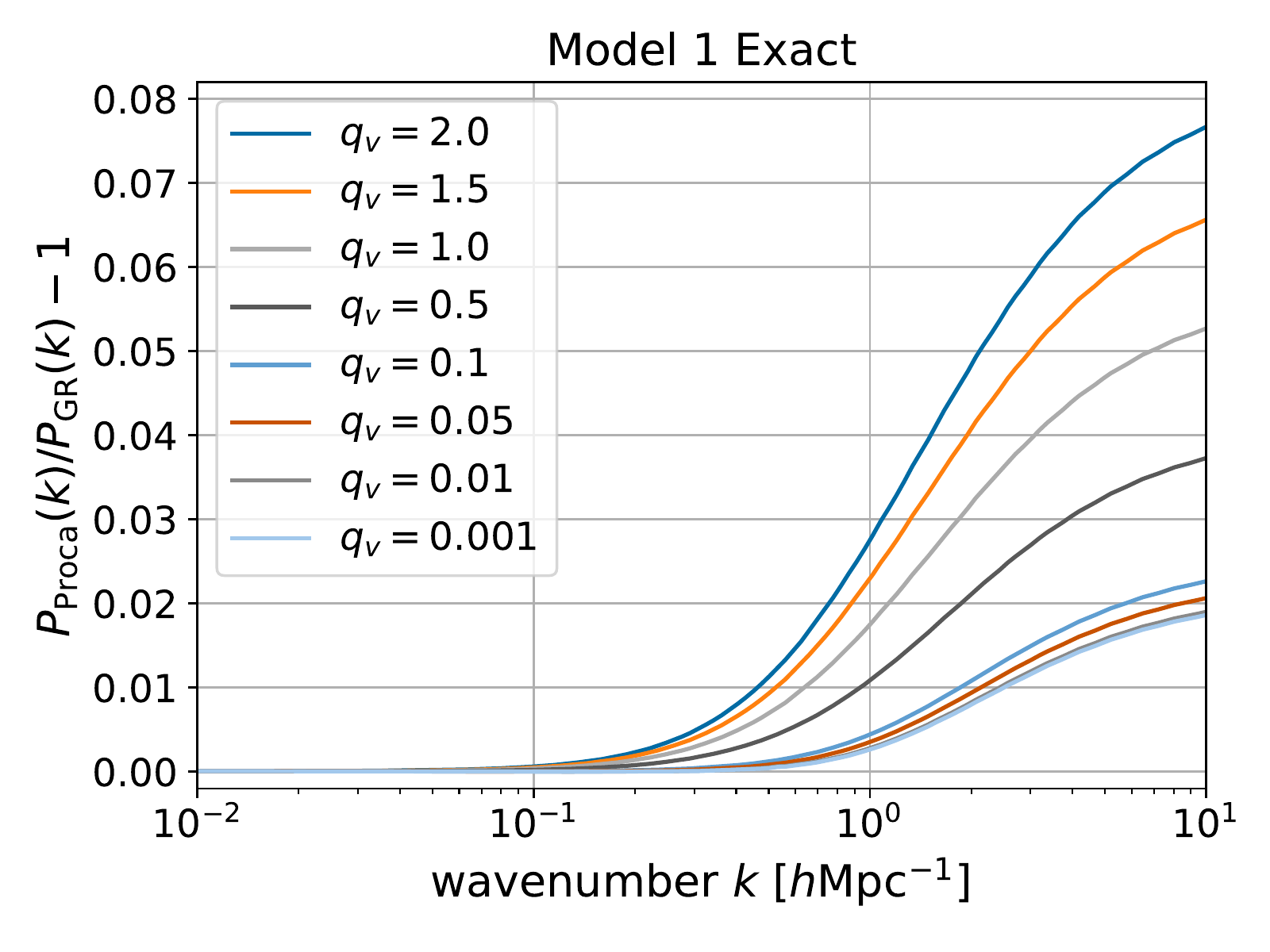}\hfill
  \includegraphics[width=0.49\hsize]{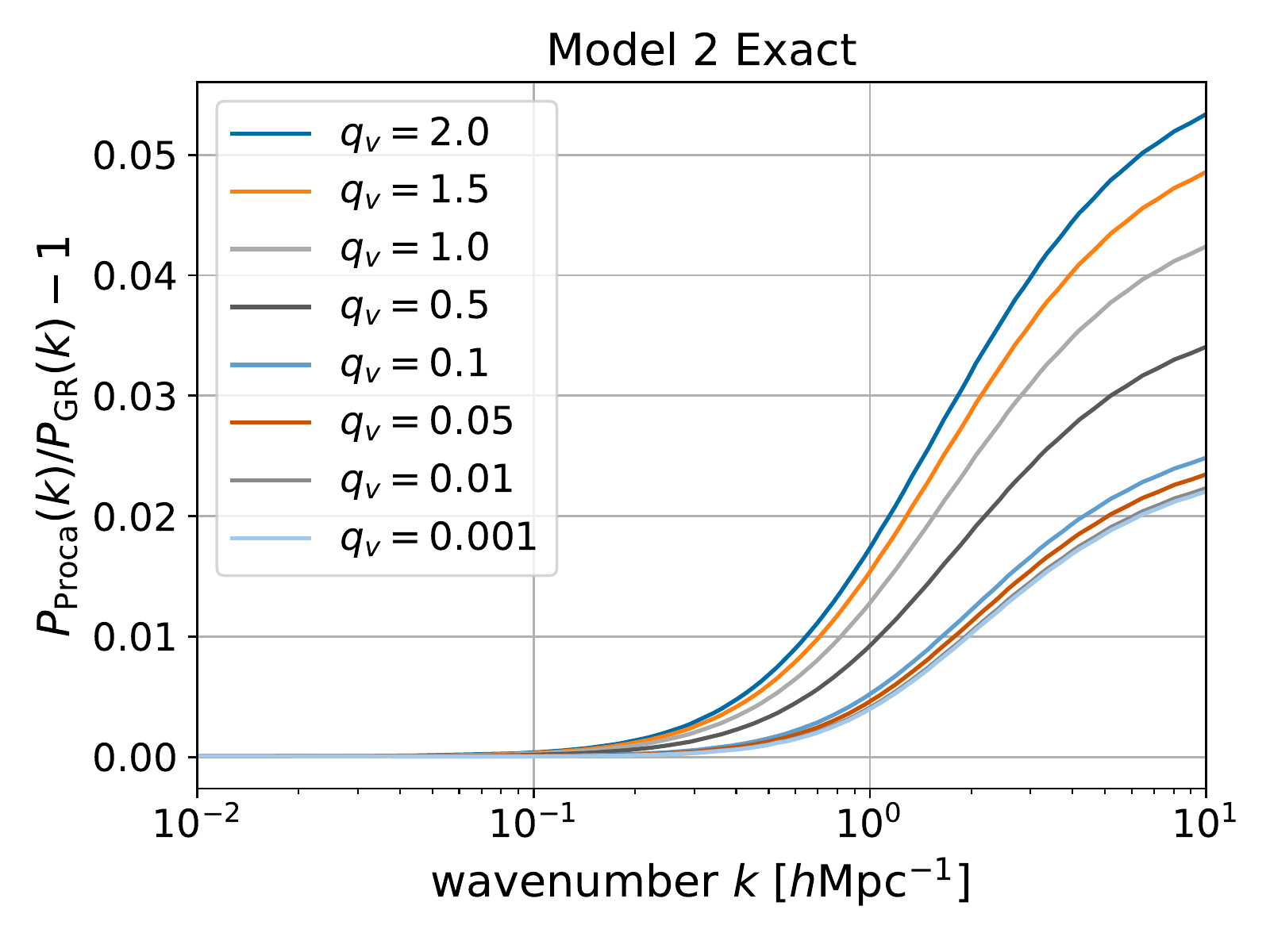}
\caption{As Fig.~\ref{fig:5}, but results obtained in a direct approach and not via the Taylor expansion.}
\label{fig:6}
\end{figure}

In order to check the validity and accuracy of our Taylor we also show results obtained by directly specifying the Proca expansion function, growth factor and gravitational constant in \eqref{eq:8}. The relative difference between this exact approach and our Taylor expansion turns out to be less than $0.15\%$ in all tested cases. The results obtained this way can be seen in Fig.~\ref{fig:6}.

The differences between Fig.~\ref{fig:5} and Figs.~2 and 3 of \cite{2019PhLB..796...59H} originate from the different normalization scheme for the initial power spectrum applied here. The shape of the curves shown in Fig.~\ref{fig:5} is generic: the wave number above which the non-linear spectrum obtained in the modified gravity theory deviates from the generally-relativistic result is set by the condition $\sigma_J \approx 1$, with $\sigma_J$ defined in (\ref{eq:5}). The amplitude of the deviations results from two competing effects illustrated in Fig.~\ref{fig:4}: the functional derivative of the non-linear power spectrum with respect to the gravitational coupling is positive, while its derivative with respect to the expansion function is negative. The quantitative consequences of these two terms for any modified gravity theory then depend on the specific changes $\Delta E$ and $\Delta G$, integrated together with these derivatives as shown in (\ref{eq:6}). Although detailed comparisons to simulations are difficult, our analytic results tend to agree well with results obtained numerically \cite[for examples]{2012JCAP...10..002B, 2013JCAP...04..029B, 2020MNRAS.497.1885H, Becker_2020, 2021JCAP...06..014B}. Comparing Figs.~\ref{fig:5} and \ref{fig:6} shows that the functional Taylor expansion applied here returns highly accurate results.

The KFT mean-field spectrum has two parameters: The scale $k_0$ where non-linear structure growth sets in, and an effective viscosity $\nu$. We have kept these parameters fixed throughout this paper, expecting them to change at most slightly such that varying them would be a second order effect. The main reason for this expectation is that the non-linearity scale in any theory must reproduce the observable scale. In principle, changing $k_0$ would modify the scale where non-linear structure growth sets in and thus also the scale where the AG power spectra would lift above or drop below the GR result. Changing $\nu$ would modify the amplitude on non-linear scales, which must also match the observations. We expect these changes to be insignificant compared to the changes included in the spectrum. Nonetheless, it would be interesting to investigate such changes in the future. To take them into account, respective functional derivatives would have to be added, and modifications would have to be constrained by observations.  We emphasize however that such extensions would not change the generality of our result: we would still see no change at the largest scales, an onset of change around the (possible slightly changed) non-linear scale and then a rising amplitude towards larger wave numbers.

\section{Comparison to Other Studies}

Changes to the non-linear density fluctuation power spectrum effected by alternative theories of gravity have also been quantified by other authors before us.

These studies are either based on $N$-body simulations, higher-order perturbation theory relative to the hydrodynamical equations or a suitable combination of the two (see for instance \cite{Ramachandra_2021, Valogiannis_2017, Tassev_2013}). While $N$-body simulations return results which are accurate within certain ranges of wave numbers set by box size and resolution, they are numerically intensive and time consuming and thus not applicable to scanning the wide landscape of alternative gravity theories developed in the literature. Furthermore, they are essentially black boxes, rendering it difficult to trace certain changes to the non-linear power spectrum back to their physical origin.  Examples of these $N$-body simulation studies are \cite{Cui_2010, Puchwein_2013, Baldi_2014,Winther_2015}.

Recently, however, some progress has also been made with analytical approaches. One approach consists of a suitable generalization of standard and Lagrangian perturbation theory to include modified gravity effects \cite[e.g.]{Koyama_2009, Aviles_2017}. Another approach is the so-called parametrized post-Friedmann framework \cite{Hu_2007}, in which non-linearities are addressed through the halo model. 

We discuss one example in more detail here, namely \cite{2019MNRAS.488.2121C}, in order to better clarify some key aspects of our setup. The authors of this paper introduce a method for calculating corrections to a $\Lambda$CDM power spectrum in a semi-analytical manner. They introduced a pseudo-cosmology in between a standard $\Lambda$CDM cosmological model and the target cosmology based on a modified gravity theory. This pseudo-cosmology is essentially a $\Lambda$CDM cosmology which already reproduces the linear clustering of the target cosmology. Using the halo model together with standard perturbation theory (SPT), they then calculated a reaction coefficient describing the deviation of the target cosmology from the pseudo-cosmology. In this way, they were able to achieve an accuracy of these analytical calculations relative to $N$-body simulations below $3\,\%$ up to wavenumbers $k\approx 1\,h\,\mathrm{Mpc}^{-1}$. This relative accuracy shrinks further below $1\,\%$ if the halo mass function as measured from simulations is included in the calculations. These results are impressive as they reach per-cent level accuracy at mildly non-linear scales. However, they rest on a number of strong assumptions and further rely on numerical simulations for the pseudo-cosmology. Using the halo model together with SPT, which becomes unreliable on fully non-linear scales due to the notorious shell-crossing problem, lets this method appear unlikely to be able to probe into scales with wave numbers $k\gtrsim 1\,h\,\mathrm{Mpc}^{-1}$. In addition, the essential recourse to $N$-body simulations renders it again difficult to trace changes in the non-linear power spectrum back to their physical origin.

With our approach, we hope to address some of these shortcomings. Using KFT allows to proceed towards much larger wave numbers since the shell-crossing problem is avoided all together by construction. The fully analytical approach of KFT further allows to identify the physical origin of changes to the non-linear power spectrum. In particular, the functional Taylor expansion reveals the reasons for characteristic changes to the power spectrum, namely modifications in the expansion function and the gravitational coupling strength, and quantifies the effects that changing these two functions has, independent of the particular modification of gravity or the cosmological model under consideration.

\section{Conclusion}

In this paper, we have pursued the following thought: The power spectrum of non-linear cosmic density fluctuations can be analytically and accurately calculated with kinetic field theory (KFT) in a mean-field approximation. At redshift $z = 0$, the KFT result agrees at the per-cent level with typical results obtained from numerical simulations for wave numbers $k\lesssim 10\,h\,\mathrm{Mpc}^{-1}$. The cosmological background model and the theory of gravity enter into this KFT result in two ways; either through the expansion function (or dimension-less Hubble function) $E$, or through a possibly time-dependent gravitational coupling strength $G$, if the small-scale gravitational potential can still be described by the Poisson equation.

For the wide class of modified gravity theories satisfying these criteria, deviations from general relativity must be small because general relativity agrees very well with empirical data on small and large scales. Then, the effect of modified gravity theories on the non-linear density-fluctuation power spectrum can be approximated by a first-order Taylor expansion of the KFT power spectrum $P_\delta^\mathrm{(nl)}$ in the mean-field approximation in terms of the functions $E$ and $G$. The functional derivatives of $P_\delta^\mathrm{(nl)}$ with respect to $E$ and $G$ are to be evaluated in the $\Lambda$CDM cosmological model based on general relativity. They are thus generic for all modified gravity theories falling into the class defined above.

We have worked out the functional derivatives of $P_\delta^\mathrm{(nl)}$ with respect to $E$ and $G$ in Sect.~3 of this paper and evaluated them in $\Lambda$CDM. In Sect.~4, we have used the results to calculate the effect of one particular modified gravity theory on non-linear cosmic structure formation, namely a generalized Proca theory \cite{2014JCAP...05..015H}, to show one example. The relative difference of the non-linear density-fluctuation power spectrum from the expectation in the $\Lambda$CDM model lifts above zero near a wave number $k\approx 1\,h\,\mathrm{Mpc}^{-1}$ defined by the filtered variance $\sigma_J$ of the linear density-fluctuation power spectrum entering into the mean-field interaction term in KFT. It reaches an amplitude of several per cent near $k\approx 10\,h\,\mathrm{Mpc}^{-1}$, in good qualitative agreement with numerical results in the literature.

We argue that this scheme of a functional Taylor expansion of the non-linear density-fluctuation power spectrum can now be used to scan wide classes of modified gravity theories. To facilitate this, we provide the functional derivatives of $P_\delta^\mathrm{(nl)}$ with respect to $E$ and $G$, evaluated in $\Lambda$CDM, in tabulated form, together with simple Python code integrating them over any deviations $\Delta E$ and $\Delta G$ that may be predicted by a modified gravity theory adapted to current observational data.

\acknowledgments 
This work was funded in part by Deutsche Forschungsgemeinschaft (DFG, German Research Foundation) under Germany’s Excellence Strategy EXC-2181/1 - 390900948 (the Heidelberg STRUCTURES Cluster of Excellence). Lavinia Heisenberg would like to acknowledge financial support from the European Research Council (ERC) under the European Unions Horizon 2020 research and innovation programme grant agreement No 801781 and by the Swiss National Science Foundation grant 179740. LH further acknowledges support from the Deutsche Forschungsgemeinschaft (DFG, German Research Foundation) under Germany’s Excellence Strategy EXC 2181/1 - 390900948 (the Heidelberg STRUCTURES Excellence Cluster).

\bibliography{main}

\end{document}